\documentclass[12pt]{article}
\usepackage[nosort]{cite}
\usepackage[usenames, dvipsnames]{xcolor}
\usepackage{graphicx}
\usepackage{multicol}
\usepackage{amsfonts}
\usepackage{amssymb}
\usepackage{amsmath}
\usepackage{afterpage}
\usepackage{setspace}
\usepackage{verbatim}
\usepackage{color}
\usepackage{longtable}
\usepackage{caption}
\usepackage{heck}
\usepackage{float}
\usepackage{slashed}
\usepackage[utf8]{inputenc}
\usepackage{braket}
\usepackage{subcaption}
\usepackage{epsfig}
\usepackage{enumerate}
\usepackage{epstopdf}
\usepackage[enableskew, vcentermath]{youngtab}
\usepackage{adjustbox}
\usepackage{multirow}
\usepackage{tikz}
\usepackage[margin=1in]{geometry}
\usepackage{titletoc}
\usepackage{hyperref}
\usepackage[percent]{overpic}
\usepackage{gensymb}
\usepackage{mathtools}
\usepackage{tikz-cd}%
\usepackage[utf8]{inputenc}
\usepackage{epsf}
\usepackage{colortbl}
\usepackage{amsmath}
\usepackage{amsfonts}
\usepackage{amssymb}
\usepackage{graphicx}
\usepackage{color}
\usepackage{psfrag}
\usepackage{cite}
\usepackage{hyperref}
\usepackage{mathrsfs}
\usepackage{tikz}
\usepackage{tikz-feynman}
\usepackage{slashed}
\usepackage{ifpdf}
\usepackage{mdframed}
\mdfdefinestyle{statementstyle}{
   frametitlerule = true,
   frametitlefont = {\normalfont\bfseries},
   frametitlebackgroundcolor = gray!10,
}
\mdtheorem[style=statementstyle]{statement}{Definition}
\providecommand{\U}[1]{\protect\rule{.1in}{.1in}}
\pdfoutput=1
\newsavebox{\mysavebox}

\hypersetup{colorlinks,citecolor=black,filecolor=black,linkcolor=black,urlcolor=black}
\usetikzlibrary{decorations.markings}

\numberwithin{equation}{section}

\usetikzlibrary{chains}
\allowdisplaybreaks
\tikzset{node distance=2em, ch/.style={circle,draw,on chain,inner sep=2pt},chj/.style={ch,join},every path/.style={shorten >=4pt,shorten <=4pt},line width=1pt,baseline=-1ex}

\newcommand{\ba}{\begin{eqnarray}}
\newcommand{\ea}{\end{eqnarray}}

\newcommand{\be}{\begin{equation}}
\newcommand{\ee}{\end{equation}}

\tikzstyle{startstop} = [rectangle, rounded corners, minimum width=3cm, minimum height=1cm,text centered, draw=black, fill=blue!10]
\tikzstyle{startstop} = [rectangle, rounded corners, minimum width=3cm, minimum height=1cm,text centered, draw=black, fill=blue!10]
\tikzstyle{io} = [trapezium, trapezium left angle=70, trapezium right angle=110, minimum width=3cm, minimum height=1cm, text centered, draw=black, fill=blue!30]
\tikzstyle{process} = [rectangle, minimum width=3cm, minimum height=1cm, text centered, draw=black, fill=orange!30]
\tikzstyle{decision} = [diamond, minimum width=3cm, minimum height=1cm, text centered, draw=black, fill=green!30]
\tikzstyle{arrow} = [thick,->,>=stealth]
\tikzset{->-/.style={decoration={
  markings,
  mark=at position #1 with {\arrow[scale=2.4]{>}}},postaction={decorate}}}
\makeatletter \@addtoreset{equation}{section} \makeatother

\renewcommand{\[}{\left[}

\newcommand{\lb}{\left(}
\newcommand{\rb}{\right)}

\colorlet{darkblue}{blue!70!black}
\colorlet{darkgreen}{green!70!black}

\hypersetup{
colorlinks=true,
linkcolor=blue,
citecolor=magenta,
}

\usepackage{tikz}
\usetikzlibrary{arrows}
\usetikzlibrary{arrows.meta}
\usetikzlibrary{shapes.geometric,calc,arrows, positioning,shapes.misc,decorations.markings}
\tikzset{
  big arrow/.style={
    decoration={markings,mark=at position 1 with {\arrow[scale=2,#1]{>}}},
    postaction={decorate},
    shorten >=0.4pt},
  big arrow/.default=black}

\pgfdeclarelayer{edgelayer}
\pgfdeclarelayer{nodelayer}
\pgfsetlayers{edgelayer,nodelayer,main}
\tikzstyle{none}=[inner sep=0pt]

\tikzstyle{NodeCross}=[draw, shape=circle, cross out, inner sep=0pt, minimum size=6pt,line width=0.25mm]
\tikzstyle{Circle}=[draw, shape=circle, black, fill=black, inner sep=0pt, minimum size=6pt]
\tikzstyle{Star}=[draw, shape=star, fill=black, star points=8, inner sep=0pt, minimum size=8pt]
\tikzstyle{CircleRed}=[draw, shape=circle, black, fill=red, inner sep=0pt, minimum size=4pt]
\tikzstyle{StarP}=[draw={rgb,255: red,128; green,0; blue,128}, shape=star, fill={rgb,256: red,128; green,0; blue,128}, star points=8, inner sep=0pt, minimum size=12pt]
\tikzstyle{ShadedCircRed}=[draw=red, shape=circle, fill={rgb, 255: red,255; green,114; blue, 118}, inner sep=0pt, minimum size=80pt, line width=0.75mm, fill opacity=0.2]
\tikzstyle{ShadedCircRed2}=[draw=red, shape=circle, fill={rgb, 255: red,255; green,114; blue, 118}, inner sep=0pt, minimum size=10pt]
\tikzstyle{ShadedCircRed3}=[draw=blue, dotted, shape=circle, fill={rgb, 255: red,255; green,114; blue, 118}, inner sep=0pt, minimum size=30pt, line width=0.6mm]
\tikzstyle{ShadedCirc}=[draw=red, shape=circle, fill={rgb, 255: red, 140; green,248; blue, 255}, inner sep=0pt, minimum size=80pt,  fill opacity=0.5]

\tikzstyle{DashedLine}=[-, densely dashed, line width=0.25mm]
\tikzstyle{DottedLine}=[-, dotted, line width=0.25mm]
\tikzstyle{ThickLine}=[-, line width=0.25mm]
\tikzstyle{ArrowLineRight}=[-, -{Stealth[scale=1.75]}, line width=0.1mm, scale=5]
\tikzstyle{ArrowLineRed}=[-, draw={rgb,255: red,191; green,0; blue,0}, -{Stealth[scale=1.75]}, line width=0.1mm, scale=5]
\tikzstyle{RedLine}=[-, draw={rgb,255: red,191; green,0; blue,0}, fill=none, line width=0.25mm]
\tikzstyle{DashedLineThin}=[-, densely dashed, line width=0.125mm, fill=none, draw=black]
\tikzstyle{DottedRed}=[-, dotted, draw={rgb,255: red,191; green,0; blue,0}, fill=none, line width=0.25mm]
\tikzstyle{DashedRed}=[-, densely dashed, draw={rgb,255: red,191; green,0; blue,0}, fill=none, line width=0.25mm]
\tikzstyle{BlueLine}=[-, draw={rgb,255: red,0; green,0; blue,191}, fill=none, line width=0.25mm]
\tikzstyle{ArrowLineBlue}=[-, draw={rgb,255: red,0; green,0; blue,191}, -{Stealth[scale=1.75]}, line width=0.1mm, scale=5]
\tikzstyle{GreenDoubleArrow}=[<->, draw={rgb,155: red,0; green,255; blue,0},  line width= 0.5mm, scale=5]
\tikzstyle{RedDoubleArrow}=[<->, draw={rgb,255: red,255; green,0; blue,0},  line width= 0.5mm, scale=5]
\tikzstyle{BlueDottedLight}=[-, dotted, draw={rgb,255: red,0; green,0; blue,255}, fill=none, line width=0.5mm]

\begin{document}

\preprint{UPR-1322-T}

\date{May 2023}

\title{Fluxbranes, Generalized Symmetries, \\[4mm] and Verlinde's Metastable Monopole}

\institution{PENN}{\centerline{$^{1}$Department of Physics and Astronomy, University of Pennsylvania, Philadelphia, PA 19104, USA}}
\institution{PENNmath}{\centerline{$^{2}$Department of Mathematics, University of Pennsylvania, Philadelphia, PA 19104, USA}}
\institution{MARIBOR}{\centerline{${}^{3}$Center for Applied Mathematics and Theoretical Physics, University of Maribor, Maribor, Slovenia}}

\authors{
Mirjam Cveti\v{c}\worksat{\PENN,\PENNmath,\MARIBOR}\footnote{e-mail: \texttt{cvetic@physics.upenn.edu}},
Jonathan J. Heckman\worksat{\PENN,\PENNmath}\footnote{e-mail: \texttt{jheckman@sas.upenn.edu}},\\[4mm]
Max H\"ubner\worksat{\PENN}\footnote{e-mail: \texttt{hmax@sas.upenn.edu}}, and
Ethan Torres\worksat{\PENN}\footnote{e-mail: \texttt{emtorres@sas.upenn.edu}}
}

\abstract{The stringy realization of generalized symmetry operators involves wrapping ``branes at infinity''.
We argue that in the case of continuous (as opposed to discrete) symmetries, the appropriate objects are fluxbranes. We use this perspective to revisit the phase structure of Verlinde's monopole, a proposed particle which is
BPS when gravity is decoupled, but is non-BPS and metastable when gravity is switched on. Geometrically,
this monopole is obtained from branes wrapped on locally stable but globally trivial
cycles of a compactification geometry. The fluxbrane picture allows us to characterize
electric (resp. magnetic) confinement (resp. screening)
in the 4D theory as a result of monopole decay. In the presence of the fluxbrane, this decay also
creates lower-dimensional fluxbranes, which in the field theory is interpreted as the creation of an additional 
topological field theory sector.}

\maketitle

\setcounter{tocdepth}{2}


\newpage

\section{Introduction}

One of the exciting recent developments in the study of quantum field theory (QFT) is the
discovery that global symmetries have an intrinsically \textit{topological} character \cite{Gaiotto:2014kfa}. The key idea in this context is that
symmetry operators are topological, and their action on local (as well as extended) operators is captured by an associated linking in the spacetime. This point of view has lead to vast generalizations of the notion of symmetries in QFTs which are collectively referred to as generalized global symmetries. There is by now an extensive and growing literature, see e.g., \cite{Cordova:2022ruw} for a review.

In the specific context of QFTs realized via string constructions, it was recently proposed that branes ``at infinity'' can produce topological symmetry operators in a localized QFT sector \cite{Apruzzi:2022rei, GarciaEtxebarria:2022vzq, Heckman:2022muc} (see also \cite{Heckman:2022xgu, Dierigl:2023jdp}). This complements the ``defect group'' perspective on higher symmetries developed earlier \cite{DelZotto:2015isa, GarciaEtxebarria:2019caf, Albertini:2020mdx, Morrison:2020ool}.

Now, an advantage of working with a string construction is that one can in principle reintroduce the effects of gravity. Indeed, once gravity is included, such global symmetries are either gauged or explicitly broken (see e.g., \cite{Banks:1988yz, Banks:2010zn, Harlow:2018tng, McNamara:2019rup}), and the specific fate of any individual symmetry likely depends on the details of a given UV completion. The paradigm of local model building provides a general approach to these issues (see e.g., \cite{Antoniadis:2000ena, Aldazabal:2000sa, Cvetic:2000st, Verlinde:2005jr, Gray:2006jb, Donagi:2008ca, Beasley:2008dc, Beasley:2008kw, Donagi:2008kj}). One first begins by engineering a QFT of interest in the limit where gravity is decoupled. This amounts to working on a non-compact ``internal'' string background with localized degrees of freedom. Switching on gravity involves embedding this construction in a globally compact background.\footnote{For recent work on generalized symmetries in the context of compact backgrounds, see e.g., \cite{Cvetic:2020kuw, Cvetic:2021sxm, Cvetic:2021sjm, Cvetic:2022uuu}.}

In this note we investigate these issues in the specific context of continuous higher-form symmetries acting on a monopole configuration introduced by Verlinde \cite{Verlinde:2006bc} in the context of QFTs engineered via D3-branes probing a local singularity of a Calabi-Yau threefold. The monopole is BPS and stable when gravity is switched off, but is non-BPS and metastable with gravity switched on.

At long distances, the QFT is a quiver gauge theory, and a heavy monopole is obtained from a D3-brane wrapping a non-compact 3-cycle in the internal geometry. There is a corresponding flux tube which attaches to this monopole, as obtained from a D3-brane wrapping a compact 2-cycle, and in the limit where gravity is switched off, the configuration is supersymmetric, and exactly stable. This can be viewed as a 5-brane wrapping a local 2-cycle in the internal geometry, and as such it produces a domain wall between a confining / deconfining phase.

Switching on gravity qualitatively changes the story. In a compact geometry, it can happen that the local 2-cycle supporting the flux tube is actually globally trivial. As studied in \cite{Verlinde:2006bc} (for a related earlier construction see \cite{Greene:1996dh}), this means the monopole is in fact metastable, and will eventually ``self-annihilate''.\footnote{In a full compactification, there may end up being additional decay channels, which can shorten the lifetime \cite{Kasai:2015exa}. Nevertheless, it is quite plausible that such decay rates can still be suppressed relative to the string scale. For related string realizations of metastable monopole configurations, see for example \cite{Aganagic:2006ex, Heckman:2007wk}.} In geometric terms, the monopole is better viewed as a puffed up 3-ball which separates a deconfined / confined vacuum. As the unwrapping occurs, the bubble expands and eventually there is a transition to a confined $U(1)^{(0)}_{\mathrm{elec}}$ gauge theory phase. Indeed, as noted in \cite{Verlinde:2006bc}, this can also be interpreted as the formation of a non-zero monopole condensate.

Our aim in this note will be to study the generalized symmetries which act on this monopole, first in the limit where gravity is switched off, and then in the limit with gravity switched on. When gravity is switched off we expand on the paradigm of the construction of generalized symmetry operators using ``branes at infinity" as presented in \cite{Apruzzi:2022rei, GarciaEtxebarria:2022vzq, Heckman:2022muc}. We argue that the correct branes to wrap ``at infinity" are fluxbranes, and we determine the worldvolume topological actions for these configurations. These naturally act on defects charged under continuous higher-form symmetries, as obtained from branes wrapped on free, relative cycles (as opposed to torsional, relative cycles). We use this perspective to track the deconfinement / confinement transition of Verlinde's configuration in an adiabatic limit where gravity is switched on. When the boundary of the 5-brane bubble collides with the fluxbrane, the Hanany-Witten effect creates a fundamental flux 2-brane. This effect parallels the creation of symmetry operators observed field theoretically in \cite{Kaidi:2021xfk, Choi:2021kmx} and from a top-down perspective in \cite{Apruzzi:2022rei, Heckman:2022xgu, Dierigl:2023jdp}.

\section{Symmetry Operators from Fluxbranes}

We begin by studying the stringy realization of generalized global symmetry operators associated with continuous higher-form symmetries, i.e., symmetry operators labeled by a continuous parameter. These act on defects engineered via branes wrapping free cycles. With this characterization in hand we then turn to an analysis of Verlinde's monopole.

Let us now consider a QFT engineered from IIA or IIB string theory on $\mathbb{R}^{D-1,1}\times X$, with non-compact internal directions $X$ of dimension $10-D$. The QFT is obtained from a localized singularity in $X$ which can involve the profile of the metric, as well as additional degrees of freedom such as spacetime filling branes.

The generalized global symmetry operators of such a QFT are constructed by wrapping branes on cycles in the asymptotic spatial boundary $\partial X$  \cite{Heckman:2022muc}. The defects furnish representations of these symmetry operators, and are constructed by wrapping $q$-branes on non-compact cycles with asymptotic boundary in $\partial X$ \cite{DelZotto:2015isa, GarciaEtxebarria:2019caf, Albertini:2020mdx, Morrison:2020ool}.

Recently a proposal was made for the corresponding symmetry operators which detect these defects \cite{Apruzzi:2022rei, GarciaEtxebarria:2022vzq, Heckman:2022muc}, which we now briefly review. At the level of the supergravity theory, there is a natural operator we can introduce which detects the corresponding field theory defect. Given a $q$-brane, observe that it couples to a $(q+1)$-form potential. There is a magnetic dual flux $F_{p+2}$ which detects such charged objects. The parameters $q$ and $p$ are related as:
\begin{equation}
(p+2) + (q+2) = 10.
\end{equation}
Observe that in the 10D spacetime, the $q$-brane naturally links with the flux operator:
\begin{equation}\label{eq:FLUXOP}
\mathcal{U}_{\eta}(Y_{p+2}) = \exp \left( 2 \pi i \eta \underset{Y_{p+2}}{\int} F_{p+2} + ... \right),
\end{equation}
where the ``...'' indicates improvement terms which must be included for self-consistency. Here, the value of the parameter $\eta$, as well as the specific type of symmetry depends on whether we are dealing with a continuous or discrete generalized symmetry. In the latter case, one can show that the symmetry operator of the $D$-dimensional QFT can be interpreted as a magnetic dual $p$-brane wrapped ``at infinity''. Geometrically, this brane links with the heavy defect both in the spacetime as well as in the boundary geometry $\partial X$. On the other hand, when the symmetry in question is continuous, we will shortly argue that the natural object in question is a fluxbrane.

To motivate this proposal, let us begin by reviewing the case of discrete / finite order symmetries. Consider wrapping a $p$-brane and the electromagnetically dual $q$-brane on
\be\begin{aligned}
M_{p+1-k}\times \gamma_k &\subset \mathbb{R}^{D-1,1}\times \partial X\,, \\
N_{q-\ell}\times \Gamma_{\ell+1} &\subset \mathbb{R}^{D-1,1}\times \partial X\,, \\
\end{aligned}\ee
respectively. We view the $q$-brane as generating a heavy defect, and as we now explain, the $p$-brane as a symmetry operator. We have $p+q=6$ and $\gamma_k\in H_k(\partial X)$ and $\Gamma_{\ell+1}\in H_{\ell+1}(X,\partial X)/H_{\ell+1}(X)$. The former constructs a generalized symmetry operator\footnote{A recurring subtlety in this procedure is that we should more properly replace $H_*$ by some generalized homology theory $\mathbb{E}_*$ which in principle classifies the possible charges of the string theory we are working in. See for instance Appendix A of \cite{Heckman:2022xgu} which reviews the appearance of twisted K-theory as the generalized (co)homology relevant for NS5 brane backgrounds. These subtleties will not be relevant for the examples in this paper.} which acts on defects of dimension $q-\ell$ constructed by the latter. The supports $M_{p+1-k}$ and $N_{q-\ell}$ link in $\mathbb{R}^{D-1,1}$ and therefore
\be
k+\ell=\mathrm{dim}(\partial X)-1
\ee
and whenever $\gamma_k, \partial \Gamma_{\ell + 1}$ are torsional cycles in $\partial X$ the canonical linking pairing
\be
L_{\partial X}:\mathrm{Tor}\,H_k(\partial X)\times \mathrm{Tor}\,H_{\ell}(\partial X)\rightarrow \mathbb{Q}/\mathbb{Z}
\ee
geometrizes the symmetry action \cite{Apruzzi:2022rei, GarciaEtxebarria:2022vzq, Heckman:2022muc}. The phase of this symmetry action is then
\begin{equation}\label{eq:linkingreview}
  \exp\left(2\pi i \times l_{MN}\times L_{\partial X}(\gamma_k,\partial \Gamma_{\ell+1})\right)
\end{equation}
where $l_{MN}$ denotes the Gauss-linking of $M_{p+1-k}$ with $N_{q-\ell}$ in $\mathbb{R}^{D-1,1}$.

Crucially, the above makes use of torsional cycles, but the general idea \cite{Apruzzi:2022rei, GarciaEtxebarria:2022vzq, Heckman:2022muc} applies more broadly. We now explain the case in which homology classes wrapped by $p$-branes and $q$-branes are free. In this case, a charged defect operator constructed from a $q$-brane wrapping $\Gamma_{\ell +1}$ is labeled by an element in a free charge lattice so we seek to understand the stringy construction of the associated (naively $U(1)$) symmetry operators.

To begin, let us return to the supergravity operator of line (\ref{eq:FLUXOP}).
Our proposal is that the symmetry operator is realized by wrapping a \textit{fluxbrane} along
\be
M_{p+1-k}\times \gamma_{k+1}\subset \mathbb{R}^{D-1,1}\times \partial X
\ee
 where $\gamma_{k+1}\in H^{\mathrm{free}}_{k+1}(\partial X)\simeq H_{k+1}(\partial X)/\mathrm{Tor}\,H_{k+1}(\partial X)$.

A fluxbrane is a higher-dimensional analog of a flux tube from QCD.\footnote{They were introduced in a four-dimensional setting in the Melvin universe \cite{Melvin:1963qx}.} In keeping with standard brane indexing conventions, we refer to a flux $(p+1)$-brane as one which couples to the curvature $F_{p+2}$ and thus occupies $p+2$ total spacetime directions. One can also construct fluxbranes as soliton backgrounds in brane / anti-brane pairs (see Appendix \ref{app:fluxfrombranes}).

Fluxbranes have been studied in string theory before (see e.g. \cite{Gutperle:2001mb,Emparan:2001gm}), and they usually break supersymmetry and are unstable against forming brane / anti-brane pairs (similar to a QCD flux tube being unstable to splitting into quark / anti-quark pairs). The dynamics of the fluxbrane is immaterial for its use as a topological symmetry operator because it is infinitely far away from the QFT degrees of freedom and wrapping the fluxbrane on a formally infinite volume cycle in $\partial X$ suppresses decay processes.\footnote{Similar observations were recently used in \cite{Dierigl:2023jdp} to study the role of various non-BPS branes as generalized symmetry operators.}

For free classes the relevant canonical pairing is now the intersection pairing
\be
(\,.\,,.\,)_{\partial X}: ~H^{\mathrm{free}}_{k+1}(\partial X)\times H^{\mathrm{free}}_{\ell}(\partial X)\rightarrow \mathbb{Z}
\ee
where $k+1+\ell=\textnormal{dim}\,\partial X$. Flux $(p+1)$-branes canonically pair with $q$-branes sourcing the flux. The supports of these two objects link in the product $\mathbb{R}^{D-1,1}\times \partial X$, rather than individually in each factor, and the leading topological term in the flux $(p+1)$-brane action is:
\begin{equation}\label{eq:fluxbraneaction1}
  \int_{M_{p+1-k}\times \gamma_{k+1}}\left( F_{p+2} +\dots  \right)
\end{equation}
which precisely measures the number of $q$-branes wrapping $N_{q-\ell}\times \Gamma_{\ell+1}$.

Self-consistency of the fluxbrane topological terms requires additional improvement terms, as indicated by the ``...'' in line \eqref{eq:fluxbraneaction1}. One way to argue for the appearance of such terms is to observe that just as we can consider a fluxbrane in isolation, it can also support lower-dimensional fluxbranes. This is simply the analog of the ``branes within branes'' observed for D-branes \cite{Douglas:1995bn}. In the present context, it is essentially \textit{forced} because we must allow for the totality of all possible stacked heavy defects generated by wrapped $q$-branes, and their corresponding symmetry operators. From a bottom up point of view, such additional terms can be argued for from a corresponding anomaly inflow analysis, and this in turn requires including additional topological terms in the action. The topological couplings we propose are essentially the minimal ones compatible with other stringy considerations.

These additional topological terms are generalizations of the Wess-Zumino terms on $p$-branes. To get a handle on them, let $\mathcal{L}^{(p+1)}_{\textnormal{WZ}}=C_{p+1}+\dots$ be the Wess-Zumino Lagrangian of a D$p$-brane.
The D$p$-brane sources the term:
\begin{equation}\label{eq:}
  \int \mathcal{L}^{(p+1)}_{\textnormal{WZ}} \wedge d*F_{p+2}= - (-1)^{p+1} \int d\mathcal{L}^{(p+1)}_{\textnormal{WZ}} \wedge * F_{p+2}
\end{equation}
in the corresponding 10D supergravity action. More precisely, we view $d\mathcal{L}^{(p+1)}_{\textnormal{WZ}}$ as the local expression for a $(p+2)$-form which we can interpret as a generalized curvature $\mathcal{F}^{(p+2)}_{\textnormal{WZ}}$. With this we also find that the topological sector of the fluxbrane supports a $U(1)$ gauge field.\footnote{It is important to stress that our considerations apply to the \textit{topological} sector of the fluxbrane. The full dynamics of a fluxbrane are more challenging to characterize, but are also irrelevant for the present analysis.} Assuming that the fluxbrane localizes on some submanifold $Y_{p+2}\subset\partial X$ in the asymptotic boundary $\partial X$, we therefore have
\be
* F_{p+2}=\eta \delta_{Y_{p+2}}
\ee
and with this the action \eqref{eq:fluxbraneaction1} is completed to
\be \label{eq:Fluxbrane}
  \int_{M_{p+1-k}\times \gamma_{k+1}}\left( F_{p+2} +\dots  \right)=\int_{M_{p+1-k}\times \gamma_{k+1}} \mathcal{F}^{(p+2)}_{\textnormal{WZ}}
\ee
and the topological operator engineered using this is
\begin{equation}\label{eq:symop}
  \mathcal{U}_{\eta}(M_{p+1-k})=\exp\left( 2\pi i \eta \int_{M_{p+1-k}\times \gamma_{k+1}} \mathcal{F}^{(p+2)}_{\textnormal{WZ}}\right).
\end{equation}
As in \cite{Apruzzi:2022rei, GarciaEtxebarria:2022vzq, Heckman:2022muc}, the terms beyond $C_{p+1}$ in $\mathcal{L}^{(p+1)}_{\textnormal{WZ}}$ enrich the fusion algebra generated by $\mathcal{U}_{\eta}(M_{p+1-k})$ from that of a $U(1)$ symmetry\footnote{The algebra for the invertible case would be the group ring $\mathbb{C}[U(1)]$.} to a non-invertible symmetry\footnote{See e.g., \cite{Komargodski:2020mxz, Choi:2021kmx, Kaidi:2021xfk}}. It may happen that $\mathcal{U}_\eta(M_{p+1-k})$ is not gauge invariant when $\eta$ is irrational unless we extend the action in \eqref{eq:symop} to some higher-dimensional manifold whose boundary is $M_{p+1-k}\times \gamma_{k+1}$. This subtlety will not play an important role in this work but we simply highlight that this is reminiscent of how non-invertible symmetries constructed from ABJ anomalies can be made gauge invariant only if a certain phase takes values in $\mathbb{Q}/\mathbb{Z}$ rather than $\mathbb{R}/\mathbb{Z}$ \cite{Choi:2022jqy,Cordova:2022ieu}.\footnote{See also the work \cite{Garcia-Valdecasas:2023mis} which shows that several examples of (exponentials of) integrals of supergravity page charges can only be made gauge invariant when the phase of the exponent is rational.} We see the same effect in Appendix \ref{app:fluxfrombranes} where the construction of fluxbranes from higher dimensional brane / anti-brane pairs only works in generality when $\eta\in \mathbb{Q}/\mathbb{Z}$. Similar remarks hold for generating fluxbrane actions from WZ terms of other types of branes in string / M-theory.

The appearance of fluxbranes as a way to engineer symmetry operators is also quite natural in the framework of differential cohomology. When defects and symmetry operators are constructed via wrapped branes on torsional classes the correct cocycle to expand over the cycle $N_{q-\ell}\times \Gamma_{\ell+1}$ is the class $\breve{\mathcal{L}}^{(p+1)}_{\textnormal{WZ}}$, the uplift of the Wess-Zumino Lagrangian to differential cohomology \cite{Apruzzi:2021nmk}. This gives a symmetry operator with
 \begin{equation}\label{eq:fluxbraneaction2}
 2 \pi i \int_{M_{p+1-k}\times \gamma_{k}}\left( \breve{F}_{p+2} +\dots  \right)
\end{equation}
in the exponent which parallels line \eqref{eq:fluxbraneaction1} and leads to an anomaly inflow formulation in the associated field theory  \cite{Apruzzi:2022rei, GarciaEtxebarria:2022vzq, Heckman:2022muc, Heckman:2022xgu}. For free classes such as $\gamma_{k+1}$, we need to integrate over $M_{p+1-k}\times \gamma_{k+1}$ in one higher dimension. Hence, it is natural to view this as a topological term for a fluxbrane in string theory.

\section{Verlinde's Monopole Revisited}

We now turn to the generalized symmetries acting on Verlinde's monopole, a metastable object which can have an exponentially long lifetime relative to the string scale, and which can have a mass which ranges from the string scale down to the $O(100)$ TeV scale depending on the details of the warping in the extra dimensions of a string compactification. We begin in the limit where gravity is switched off, and then turn to the implications of switching on gravity.

To frame the discussion to follow, we now briefly review Verlinde's monopole configuration. We engineer a 4D gauge theory on $\mathbb{R}^{3,1}$ by considering a stack of $N$ D3-branes probing $X$, a local Calabi-Yau singularity. The main idea is that the probe D3-brane ``fractionates'' and is instead replaced by various bound states of higher-dimensional branes and anti-branes wrapping 2-cycles and 4-cycles in a resolution $\widetilde{X}$ \cite{Douglas:1996sw, Diaconescu:1997br, Hori:2000ck}. These states are nevertheless mutually supersymmetric in the regime of small volume. By working at strong string coupling, one can also entertain various F-theory models, as obtained from intersecting 7-branes wrapping various 4-cycles of the geometry. In this case, one does not consider the collapsing cycle limit, and moreover, one also relaxes the Calabi-Yau condition on $X$.\footnote{Rather, the topological twist on the 7-branes leads to an effective local Calabi-Yau geometry (see \cite{Bershadsky:1997zs, Beasley:2008dc}).} For ease of exposition, we focus on the case of probe D3-branes, and also assume that $\widetilde{X}$ is given by $K \rightarrow S$, the canonical bundle of a single K\"ahler surface $S$ which can be contracted to a point (i.e., it is Fano). That said, the considerations we present generalize to many other settings.

Coupling to gravity is accomplished by viewing $X$ as a local patch of a compact Calabi-Yau $Y$. The value of Newton's constant (in Einstein frame) is then set by $G_{\mathrm{Newton}} \sim 1 / \mathrm{Vol}(Y)$. The specific UV completion clearly depends on these details, explicit examples in D3-brane probe / F-theory models include those of references \cite{Buican:2006sn, Beasley:2008kw, Donagi:2008kj}.

We are interested in models with a $U(1)^{(0)}_{\mathrm{elec}}$ gauge symmetry which we refer to as ``hypercharge'' (since this is where it often shows up in this context), though clearly one can entertain more general models. The subscript ``elec'' denotes the fact that we work in a global realization of the theory in which the corresponding electric degrees of freedom are light, and in which the corresponding Wilson lines are part of the spectrum of line operators. When there is no confusion, we shall suppress this subscript.

In both local D3-brane probe models, as well as F-theory models, the condition that this $U(1)$ remains massless requires a specific geometric condition be met, namely that there is a 2-cycle $\alpha \subset H_{2}(S)$ which is non-trivial in the local resolution $\widetilde{X}$, but which is trivial in $Y$.\footnote{A simple example for F-theory models is to take $S = \mathbb{P}^1 \times \mathbb{P}^1$ and $Y = \mathbb{P}^3$. Letting $\sigma_1, \sigma_2$ denote the $\mathbb{P}^1$ classes of $H_{2}(S)$, observe that $\alpha = \sigma_1 - \sigma_2$ is trivial in $Y$. This follows trivially from the fact that the homology ring of $\mathbb{P}^3$ is generated by the hyperplane class, and $\mathbb{P}^1 \times \mathbb{P}^1$ is specified by a quadric.} Doing so ensures that possible couplings of bulk RR-forms to local field strengths are absent, thus preventing a mass via the St\"uckelberg mechanism (see e.g., \cite{Blumenhagen:2005mu} for a review of this issue).

The essential idea in \cite{Verlinde:2006bc} is to now construct a metastable configuration by exploiting the triviality of $\alpha$ in the full geometry $Y$. From the perspective of the 4D gauge theory, we consider a $U(1)$ monopole configuration, as obtained by letting $Q$ units of $F_{\mathrm{hyp}}$ flux thread an $S^2$ in the spacetime. Topologically, we introduce a 3-ball $B$ of radius $R$, and split the spatial $\mathbb{R}^{3}$ as $\mathbb{R}^{3} \backslash B$ and $B$. Integrating $F_{\mathrm{hyp}}$ over this $S^2$ specifies the total charge:
\begin{equation}
\underset{B}{\int} d F_{\mathrm{hyp}} = \underset{S^2}{\int} F_{\mathrm{hyp}} = Q.
\end{equation}
The ``core size'' of the monopole is specified by the radius of the 3-ball $B$. In the context of a string construction, there is a natural (typically string scale) size for this object based on balancing the internal and external tensions from wrapped branes.

\begin{figure}[t!]
\centering
\includegraphics[scale=0.45, trim = {0cm 3.0cm 0cm 3.0cm}]{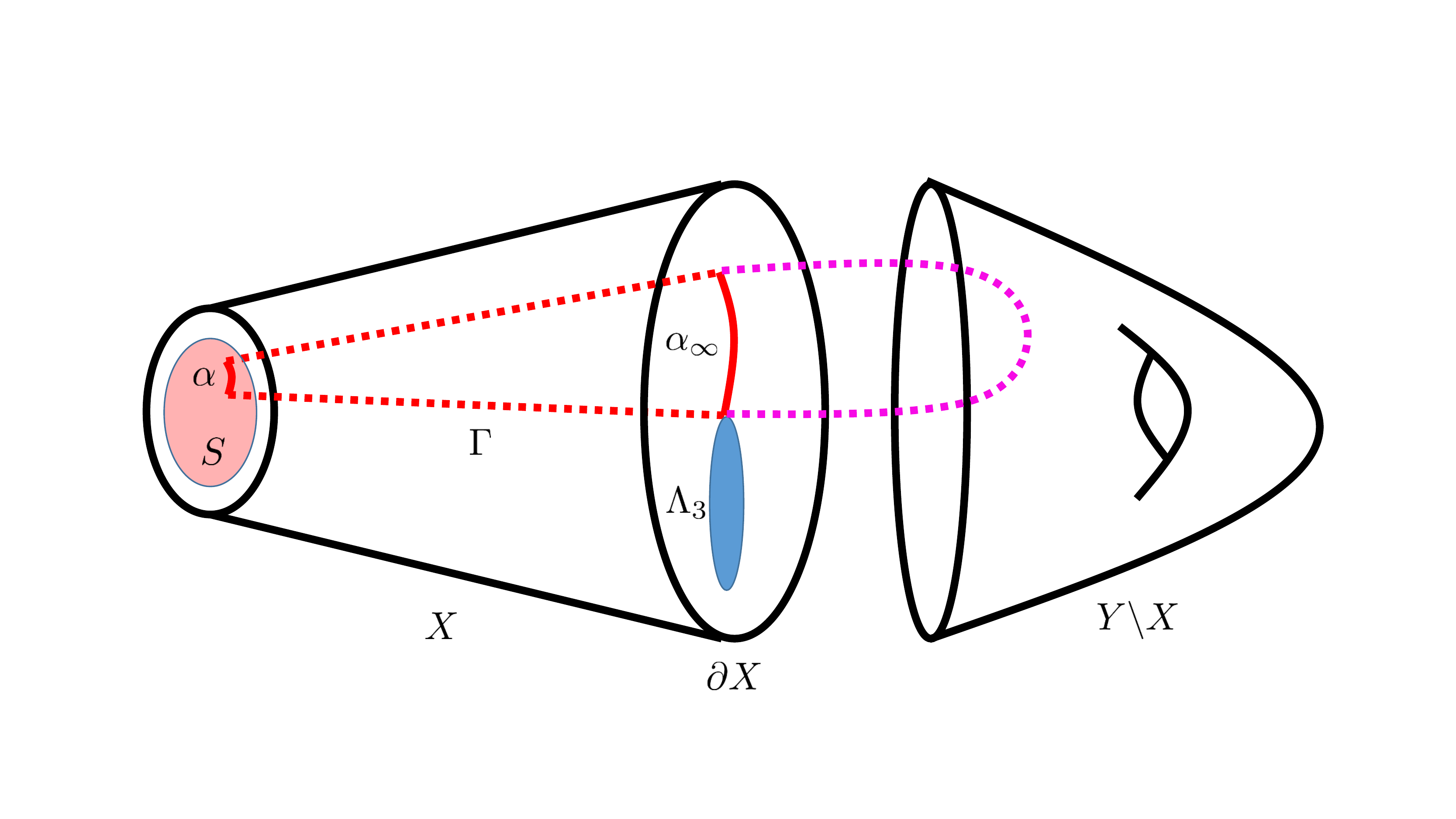}
\caption{Depiction of the local geometry used to engineer the metastable monopole. Locally, we have $X$ as given by the canonical bundle over a K\"ahler surface $S$. In this local surface $S$, we have a locally stable 2-cycle $\alpha$ which is the boundary of a non-compact 3-chain $\Gamma$ which extends along the radial direction of $X$ to the boundary $\partial X$, where the image 2-cycle (under the Gysin sequence) is denoted as $\alpha_{\infty}$. In the full compact geometry $Y$, this 3-chain can unwind, so the resulting defect is only metastable. We have also indicated the 3-cycle $\Lambda_3$ which intersects $\Gamma$ at a point in $\partial X$. The internal cycle $\Lambda_3$, when integrated against a 5-form topological term of the fluxbrane results in a 2D TFT in the spacetime which detects the monopole defect.}
\label{fig:LocGlob}
\end{figure}

Indeed, from the perspective of the internal geometry, we consider spacetime filling D5-branes wrapped on the cycle $\alpha$.\footnote{In the explicit hypercharge model of \cite{Verlinde:2006bc}, this is typically a bound state of branes and anti-branes, but this complication plays no significant role in what follows.} In the D5-brane worldvolume, there is a topological coupling $F_{\mathrm{hyp}} \wedge C_4$, and so switching on $F_{\mathrm{hyp}}$ with $Q$ units of flux can be interpreted as introducing $Q$ D3-branes wrapping $\alpha$, and extending as a one-dimensional effective string in the spacetime. The end of the string is a heavy monopole, and this can be interpreted as a D3-brane wrapping a 3-chain $\Gamma$ with boundary $\partial \Gamma = \alpha$. This 3-chain extends from the tip of the local geometry out to ``infinity''. See figure \ref{fig:LocGlob} for a depiction of the internal local / global geometry.

\subsection{Defects and Symmetry Operators}

Since the 3-chain is non-compact, the resulting monopole is really a heavy defect / line operator. This is in line with the general stringy picture of engineering defects from wrapping branes on non-compact cycles \cite{DelZotto:2015isa, GarciaEtxebarria:2019caf, Albertini:2020mdx, Morrison:2020ool}, and just as there, we can study the states which cannot be screened by dynamical states. Compared with the main examples studied previously, our main focus will be on factors of the defect group which are not torsion, i.e., they involve copies of $\mathbb{Z}$, and so the Pontryagin dual defining the higher symmetry is a continuous $U(1)$.

Given a 1-form symmetry $U(1)^{(1)}_{\mathrm{mag}}$, there is a corresponding topological symmetry operator which we can obtain by integrating the RR 5-form flux over a 5-cycle which links with the 4-cycle wrapped by the D3-brane on $\mathbb{R}_{\mathrm{time}} \times \Gamma$. This constructs a flux operator which measures the D3-brane flux threaded through the 5-cycle \cite{GarciaEtxebarria:2019caf, Apruzzi:2022rei, GarciaEtxebarria:2022vzq, Heckman:2022muc,Heckman:2022xgu}. The string construction suggests that we should complete this operator into a fluxbrane as introduced in \eqref{eq:Fluxbrane}. Concretely we have in the context of Verlinde's monopole
\begin{equation}\label{eq:SYMMOP}
\mathcal{U}_\eta(\Sigma_2) = \exp \left(2 \pi i \eta \underset{\Sigma_2 \times \Lambda_3}{\int} F_5 + ...\right),
\end{equation}
where we integrate over a 3-cycle $\Lambda_3$ in $\partial X$ which intersects $\Gamma$ out in $\partial X$, and where $\Sigma_2$ links with the monopole line in the 4D spacetime. See the lefthand side of figure \ref{fig:LocGlob}.

The local profile for the heavy monopole, effective string, and symmetry operator are obtained from filling the following directions in the 10D spacetime:
\begin{equation}%
\label{table}
\begin{tabular}
[c]{|c|c|c|c|c|c|c|c|c|c|c|}\hline
& $0$ & $1$ & $2$ & $3$ &   $4$ &   $5$ & $6$ & $7$ & $8$ & $9$\\\hline
Mono Defect & $\times$ &  &  &  &   $\times$   & $\times$ & $\times$ &  &  &
\\\hline
Mono String & $\times$ & $\times$ &  &  &  0     & $\times$ & $\times$ &  &  &
\\\hline
Symm Op &  &  & $\times$ & $\times$ &    $\infty$   &  &  & $\times$ & $\times$ &
$\times$\\\hline
\end{tabular}
\end{equation}
where the directions ``$0,1,2,3$'' are in the 4D spacetime, ``$4$'' is the radial direction of $X$ (viewed as a cone over $\partial X$) and ``$5,6,7,8,9$'' are directions of $\partial X$. The string terminated by the monopole wraps cycles in $S$, the zero section of $K\rightarrow S$ and as such they sit at $r=0$. The symmetry operator sits at $r=\infty$ and the heavy defect fills the radial direction.

Summarizing, we have the fluxbrane symmetry operator (we leave the overall normalization in the path integral implicit):
\begin{equation}\begin{aligned} \label{eq:SYMMOPFULL}
\mathcal{U}_\eta(\Sigma_2) &= \int DA_1 \exp \left(2 \pi i \eta \underset{\Sigma_2 \times \Lambda_3}{\int}
{F}_{5} +{F}_{3}\cup\mathcal{{F}}_{2}+\frac
{1}{2}{F}_{1}\cup  \mathcal{{F}}_{2}\cup\mathcal{{F}}_{2} \right)\,,
\end{aligned}
\end{equation}
labelled by $\eta \in \mathbb{R}/\mathbb{Z}$ where $\mathcal{F}_2=F_2-B_2$ and $F_2$ is the field strength of the worldvolume gauge field $A_1$. The $F_i$ with odd $i$ denote the (pullback to the worldvolume) field strengths of the RR supergravity background fields. We have dropped a term involving the Euler class of $\Sigma_2\times \Lambda_3$.

To determine the structure of the symmetry operator in the 4D QFT, we now turn to the reduction of our topological terms on the cycle $\Lambda_{3}$. To this end, we first study in more detail the geometry of local 2-cycles in $S$ and how they lift to cycles in $\partial X$.

\subsection{Dimensional Reduction}

To carry out the dimensional reduction of our symmetry operator, let us consider the geometry of the setup more closely. Since the non-compact Calabi-Yau $X$ is just the canonical bundle $K \rightarrow S$, the boundary geometry is a circle bundle
\be
S^1~\hookrightarrow~ \partial X ~\rightarrow~ S\,, \qquad F_S=[K]
\ee
whose curvature is the canonical class. As such, one can use the Gysin sequence to track how local 2-cycles in $S$ embed in $\partial X$, as well as how they lift to 3-cycles upon inclusion of the $S^1$ bundle direction. With this there is an image of $\alpha$ ``at infinity'' which intersects $\Lambda_3$:
\begin{equation}\label{eq:Int}
\Lambda_3 \cap_{\:\!\partial X} \alpha_{\infty} = 1,
\end{equation}
in the obvious notation. Indeed, $\Lambda_3$ can be represented by the circle bundle of a 2-cycle $\beta$ of $S$. The main condition we need to ensure is that $\beta$ intersects $\alpha$ in $S$.

In more detail, the Gysin sequence
\be
\dots ~\rightarrow~ H^n(\partial X)~\rightarrow~H^{n-1}(S)~\xrightarrow[]{~F_S\,\wedge~}~H^{n+1}(S)~\rightarrow~H^{n+1}(\partial X)
~\rightarrow~\dots
\ee
produces two exact subsequences, cut out by $H_1(S)=H_3(S)=0$,\footnote{Recall that here we assume $S$ is Fano.}
relating cocycles of the base with those of the total space
\be \label{eq:ES} \begin{aligned}
0 ~\rightarrow~ H^1(\partial X)~\rightarrow~&H^{0}(S)~\xrightarrow[]{~F_S\,\wedge~}~H^{2}(S)~\rightarrow~H^{2}(\partial X)
~\rightarrow~0 \\
0 ~\rightarrow~ H^3(\partial X)~\rightarrow~&H^{2}(S)~\xrightarrow[]{~F_S\,\wedge~}~H^{4}(S)~\rightarrow~H^{4}(\partial X)
~\rightarrow~0\,.
\end{aligned}
\ee
With this we have the relations between base and boundary cohomology
\be
H^2(\partial X)=H^2(S)/\langle F_S\rangle \,, \qquad H^3(\partial X)=H^2(S)|_{\perp \;\!F_S}\,.
\ee
The spaces $\partial X$ and $S$ are smooth and by Poincar\'e duality we obtain an identification of 2-cycles and 3-cycles in $\partial X$ with (equivalences classes of) curves in $S$.

In particular, 3-cycles in $H_3(\partial X)\cong H^2(\partial X)$ have representatives induced from an $S^1$ fibration over a representative of the corresponding curve. The intersection pairing
\be
\kappa_{\perp}\,:~H_2(\partial X)\times H_3(\partial X)\rightarrow \mathbb{Z}
\ee
used in \eqref{eq:Int}, is therefore induced from $\kappa$, the intersection form on $S$. On cohomology we have an induced pairing with components $\kappa_{ij} = \sigma_i \cdot \sigma_{j}$. Here, the $\sigma_{i}$ are a basis for $H^2(S)\cong\mathbb{Z}^r$ and $r$ denotes the rank of the curve lattice of $S$.

Let us evaluate the exact subsequences in \eqref{eq:ES}. The maps marked in \eqref{eq:ES} derive from the intersection form $\kappa_{ij}$ and we can expand the curvature of the circle bundle as:
\be
F_S=\sum_{i=1}^{r} n_{S,i} \sigma_i
\ee
with integers $n_{S,i}\in \mathbb{Z}$. With this the two non-trivial maps above are
\be \label{eq:SES} \begin{aligned}
F_S\,\wedge\,:& ~ \mathbb{Z}\cong H^0(S)\rightarrow H^2(S) \cong \mathbb{Z}^r\,, \quad 1 \mapsto F_S \\
F_S\,\wedge\,:& ~ \mathbb{Z}^r\cong H^2(S)\rightarrow H^4(S) \cong \mathbb{Z}\,, \quad \sigma = m_j\sigma_j \mapsto m_j\sigma_j\wedge F_S=n_{S,i}m_j\kappa_{ij}\,.
\end{aligned} \ee
We define
$g=\underset{i}{\textnormal{gcd}}(n_{S,i})$ and find the cohomology groups
\be
H^n(\partial X)=\{ \mathbb{Z},0,\mathbb{Z}^{r-1}\oplus \mathbb{Z}_g,\mathbb{Z}^{r-1},\mathbb{Z}_g,\mathbb{Z}\}
\ee
which via the universal coefficient theorem imply the homology groups
\be
H_n(\partial X)=\{ \mathbb{Z},\mathbb{Z}_g,\mathbb{Z}^{r-1},\mathbb{Z}^{r-1}\oplus \mathbb{Z}_g,0,\mathbb{Z}\}\,.
\ee
In transitioning from cohomology to homology the arrows in \eqref{eq:SES} are reversed, and $n$-cycles of $\partial X$ map onto $n$-cycles of $S$. So, we now have maps $H_n(\partial X)\rightarrow H_n( X)\cong H_n(S)$ induced from the embedding $\partial X\rightarrow X$ lifted to degree $n$ in homology combined with deformation retraction along the radial direction to $S$.

Returning to our fluxbrane action, observe that $\Lambda_3$ is dual to the 2-form $u_{2,\Lambda}$ which is such that
\begin{equation}\begin{aligned} \label{eq:SYMMOPFULL2}
\mathcal{U}_\eta(\Sigma_2) &= \int DA_1 \exp \left( 2 \pi i \eta \underset{\Sigma_2 \times \partial X}{\int}
  u_{2,\Lambda} \cup \left( {F}_{5} +{F}_{3}\cup\mathcal{{F}}_{2}+\frac
{1}{2}{F}_{1}\cup  \mathcal{{F}}_{2}\cup\mathcal{{F}}_{2}\right) \right)\,.
\end{aligned}
\end{equation}

We are now in position to perform the integral over the asymptotic boundary.
The integral cohomology ring of $\partial X$ is generated by
\be
{1}, {u}_{2,n}, {t}_{2},{u}_{3,n}, {t}_{4}, {\textnormal{v}}\textnormal{ol} _{\,\partial X}
\ee
where ${1}$ is the degree zero cocycles and ${\textnormal{v}}\textnormal{ol} _{\,\partial X}$ the degree five cocycle. Torsional, respectively free, classes of degree $k$ are denoted by ${t}_k$, and ${u}_{k,n}$, where $n=1,\dots, r-1$. The integral over $\partial X$ now proceeds via the expansions
\be \begin{aligned}
{F}_{5}&={G}_{5}^{(5)}\cup 1 + {C}_{3}^{(5)}\cup {t}_2+ {G}_{3,n}^{(5)}\cup {u}_{2,n}+ {G}_{2,n}^{(5)}\cup {u}_{3,n}+{C}_{1}^{(5)}\cup {t}_4\\
{F}_{3}&={G}_{3}^{(3)}\cup 1 + {C}_{1}^{(3)}\cup {t}_2+ {G}_{1,n}^{(3)}\cup {u}_{2,n}+ {G}_{0,n}^{(3)}\cup {u}_{3,n}\\
{F}_{1}&={G}_{1}^{(1)}\cup 1 \\
{\mathcal F}_{2}&={\mathcal{H}}_{2}^{(2)}\cup 1+ {\mathcal{D}}_{0}^{(2)}\cup {t}_2+ {\mathcal{H}}_{0}^{(2)}\cup {u}_2
\end{aligned}\ee
with which we find
\begin{equation}\begin{aligned} \label{eq:SYMMOPFULL3}
\mathcal{U}_\eta (\Sigma_2) &= \int DA_1 \exp \left(2 \pi i\eta \lambda_n  \underset{  \Sigma_2}{\int} \lb
{G}_{2,n}^{(5)} +{G}_{0,n\,}^{(3)} \mathcal{{F}}_{2}\right) \right), \qquad
\lambda_n= \underset{  \partial X}{\int} {u}_{2,\Lambda}\cup {u}_{3,n}\in \mathbb{Z} \,.
\end{aligned}
\end{equation}
If we denote by $\beta_n\in H^2(S)|_{\perp \;\!F_S} \cong H^3(\partial X)$ the 2-cocycles corresponding to $u_{3,n}$ and $[\delta_{2}]\in  H^2(S)/\langle F_S \rangle  \cong H^2(\partial X)$ the equivalence class of 2-cocycles corresponding to $u_{2,\Lambda}$, then  we can compute the intersection number $\lambda_n = \delta_2\cdot \beta_n$ as an intersection in the surface $S$. We can now simplify further to
\be
\mathcal{U}_\eta (\Sigma_2) = \mathcal{U}^{\textnormal{(inv)}}_\eta(\Sigma_2)\,\mathcal{U}_\eta^{\textnormal{(cond)}}(\Sigma_2),
\ee
where we introduced the shorthand notation
\begin{equation}\begin{aligned} \label{eq:SYMMOPFULL4}
\mathcal{U}^{\textnormal{(inv)}}_\eta(\Sigma_2) &= \exp \left(2 \pi i \eta \lambda_n  \underset{  \Sigma_2}{\int}
{G}_{2,n}^{(5)} - {G}_{0,n\,}^{(3)} {{B}}_{2} \rb \\
\mathcal{U}^{\textnormal{(cond)}}_\eta(\Sigma_2) &= \int DA_1 \exp \left(2 \pi i \eta \lambda_n  \underset{  \Sigma_2}{\int} {G}_{0,n\,}^{(3)} {{F}}_{2} \right)
\end{aligned}
\end{equation}
and $\mathcal{U}^{\textnormal{(cond)}}_\eta$ is a condensation operator\footnote{See e.g., \cite{Gaiotto:2019xmp}.} from a 2-gauging of a 3-form symmetry\footnote{See \cite{Roumpedakis:2022aik}.} with background field $G^{(3)}_{0,n}$. The condensation operator is a projection operator $(\mathcal{U}_\eta^{\textnormal{(cond)}})^2=\mathcal{U}_\eta^{\textnormal{(cond)}}$ (properly normalized) and trivially non-invertible on its kernel.

\subsection{TFT Creation in the Finite Size Limit \label{ssec:FINITE}}

Summarizing our discussion up to this point, we have considered some of the defects obtained in a class of QFTs with gravity decoupled. In this limit, Verlinde's monopole is BPS, and we have argued that a flux 4-brane wrapped on a three-cycle is the natural object which realizes the corresponding topological symmetry operator. In the conformal limit where all mass scales are either zero or infinite, the monopole is formally infinite in mass, and has a suitable delta function support. Blowing up the collapsed two-cycles induces various mass scales in the QFT, and allows the monopole to pick up a finite mass and non-zero Compton wavelength. In this case the monopole has a core size, and so we can speak of an ``inside'' and ``outside'' to the monopole configuration. Outside, everything is just as before; the symmetry operator engineered by the fluxbrane detects the 1-form symmetry associated with the monopole line. Inside, however, we can contract the symmetry operator to a point. This is just an indication that we have broken the magnetic 1-form symmetry and monopole condensation has occurred, namely electric degrees of freedom are now confined.

The contraction of the symmetry operator inside the core of the monopole is not entirely trivial, however. As the flux 4-brane passes through the D5-brane, the linking configuration of the two branes changes and eventually they intersect along a spatial 2-sphere $\Sigma_2$ before the D5-brane encloses the symmetry operator. The flux 4-brane supports a $U(1)$ gauge field with field strength $F_2$ and by construction couples to the bulk $B_2$ field such that $F_2-B_2$ is gauge invariant, that is to say, $F_2-B_2$ is a globally well-defined 2-form on the flux 4-brane worldvolume. Similarly we have that the electromagnetic dual $*F_2-C_2$ is a globally well-defined 2-form on the flux 4-brane. Fundamentally, this observation is the starting point for the famous Hanany-Witten brane creation effect \cite{Hanany:1996ie} and which we now find to equally well apply to fluxbrane creation.\footnote{The relevance of Hanany-Witten moves in brane creation of condensation defects was noted in \cite{Apruzzi:2022rei} and has also been recently explored in \cite{Heckman:2022xgu, Dierigl:2023jdp}.}
Generally, when fluxbranes pass through D-branes a fluxbrane can be created.

Let us consider this fluxbrane creation process for our D5-brane and flux 4-brane with worldvolume $W_{\textnormal{D5}}=\Sigma_2\times \mathbb{R}_t\times \Gamma$ and $W_{\textnormal{F4}}=\Sigma_2\times \Lambda_3$ respectively. Here the spatial sphere $\Sigma_2\subset \mathbb{R}^3$, albeit with distinct radii, is shared and the four-manifold $\mathbb{R}_t\times \Gamma$ and three-manifold $\Lambda_3$ are disjoint and contained in the eight-manifold $M_8=\mathbb{R}_t\times \mathbb{R}_{\geq 0}\times X$. With this the supports of the D5-brane and flux 4-brane in $M_8$ have the correct dimensionality to link in $M_8$. Here $\mathbb{R}_{\geq 0}$ parameterizes the spatial radius of the $\mathbb{R}^3$ appearing in the 10D setup $\mathbb{R}_t\times \mathbb{R}^3\times X$, see also \eqref{table}.

Now, define the topological linking invariant
\be
L(W_{\textnormal{D5}},W_{\textnormal{F4}})=\int_{W_{\textnormal{F4}}} \frac{F_3^{\textnormal{RR}}}{2\pi}
\ee
measuring the 3-form flux $F_3^{\textnormal{RR}}$ sourced by the D5-brane through $W_{\textnormal{F4}}$. We have
\be
\int_{W_{\textnormal{F4}}} \frac{F_3^{\textnormal{RR}}}{2\pi}=\int_{W_{\textnormal{F4}}} \frac{d (C_2-*F_2+*F_2)}{2\pi}=\int_{W_{\textnormal{F4}}} \frac{d *F_2}{2\pi}
\ee
where $F_2$ is the $U(1)$ field strength of the flux 4-brane. We apply Stokes' theorem and use the fact that $C_2-*F_2$ is a 2-form globally defined on the flux 4-brane. As in the original Hanany-Witten effect we conclude that sources for $d*F_2$ are created whenever the supports $W_{\textnormal{D5}}, W_{\textnormal{F4}}$ are moved across each other such that their linking $L(W_{\textnormal{D5}},W_{\textnormal{F4}})$ changes. These have the interpretation of fundamental flux 2-branes (the fluxbranes associated with fundamental strings) stretching between the D5-brane and flux 4-brane. The 3D worldvolume of the 2-flux brane is $\Sigma_2\times I$ where $I\subset \mathbb{R}_{\geq 0}$ is some spatial, radially running interval with endpoints on the D5-brane and flux 4-brane. The action of the created fundamental flux 2-brane is
\be
-2 \pi i \eta \int_{\Sigma_2\times I} H_3\,.
\ee
This sort of finite size effect is inevitable when coupling to gravity,
see figure \ref{fig:Flux2} for a depiction of the related symmetry operator manipulations in this case.

\subsection{Switching on Gravity}

\begin{figure}[t!]
\centering
\includegraphics[scale=0.45, trim = {0cm 3.0cm 0cm 3.0cm}]{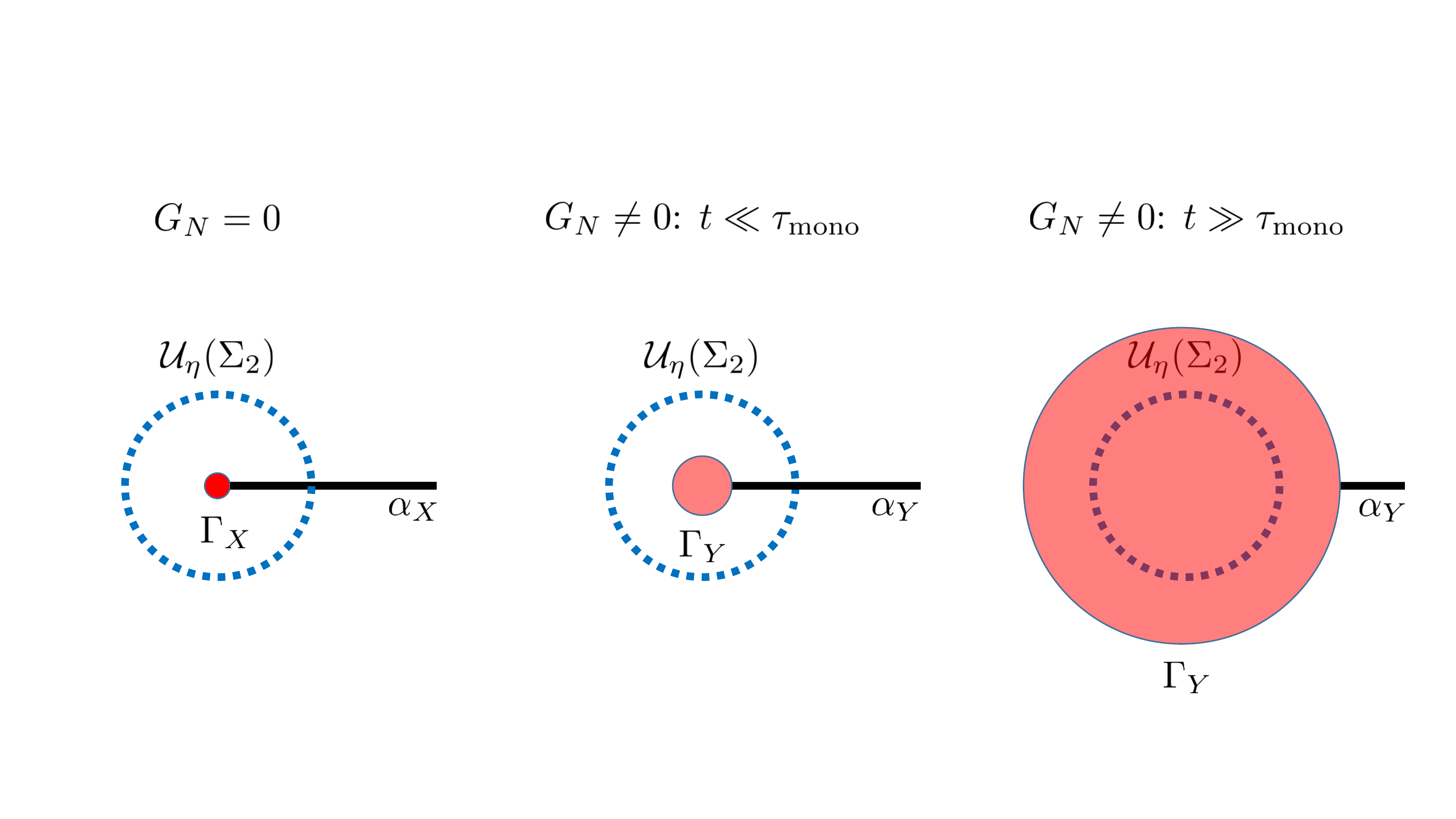}
\caption{Depiction of the monopole / flux tube configuration before and after switching on gravity, as indicated by the value of Newton's constant $G_N$ in the 4D model. In the limit where gravity is switched off (left), we have a heavy line operator as obtained from a D3-brane wrapping a non-compact 3-chain $\Gamma$, and a stringlike flux tube obtained from wrapping a D3-brane on the compact 2-cycle $\alpha$. There is a corresponding symmetry operator obtained from integrating a 5-form over an internal 3-cycle, resulting in a codimension-two topological 1-form symmetry operator $\mathcal{U}_\eta(\Sigma_2)$ which links with the heavy monopole line operator.  When gravity is switched on (middle and right), the infinitesimally small 3-ball defining the monopole begins to expand, which in the internal geometry signals the unwinding of $\alpha$ in the full compact geometry $Y$. At early times $t \ll \tau_{\mathrm{mono}}$ below the lifetime of the monopole, this 3-ball is still surrounded by $\Sigma_2$, but at late times $t \gg \tau_{\mathrm{mono}}$ above the lifetime of the monopole, the ball has expanded, and the symmetry operator no longer surrounds a monopole. In this limit, the $U(1)^{(1)}_{\mathrm{mag}}$ is broken, and the electric degrees of freedom have become confined.}
\label{fig:MonoDefect}
\end{figure}

Suppose that we now switch on gravity. In general, this will depend on how we glue the local geometry $X$ into a compact model $Y$. This will in turn affect the structure of the ensuing dynamics. One possibility is that the 2-cycle $\alpha$ embeds non-trivially in $H_2(Y)$. In this case, there is a stable object associated with the corresponding flux tube.

The case of interest to us here will be the opposite situation where $\alpha$ is actually trivial.\footnote{Geometries with this property are for example discussed in \cite{Buican:2006sn} Concretely one can consider a degree $18$ hypersurface in the weighted projective space $\mathbb{WP}(1,1,1,6,9)$ with singularities that resolve to del Pezzo surfaces. See \cite{Balasubramanian:2005zx} for more details on this particular example. For F-theory related examples, see e.g., \cite{Beasley:2008kw, Donagi:2008kj} as well as \cite{Marsano:2009ym}.} The monopole configuration we have been describing is actually metastable: While $\alpha$ is locally a minimal size 2-cycle, globally in $Y$ it can annihilate. As argued in \cite{Verlinde:2006bc}, the decay rate depends on the size of the 3-chain $\Gamma$ used to unwind the cycle.\footnote{As mentioned in \cite{Verlinde:2006bc} we must necessarily turn on a flux $\int_{\Gamma}H_3=1$ so that as the D5 wrapping $\alpha$ unwinds along $\Gamma$, the coupling $\int_{D5} B_2\wedge C_4$ makes the final D3 charge zero in the confined phase.} While this depends on many model dependent factors it should immediately be clear that this decay rate is exponentially suppressed relative to the string scale.\footnote{In \cite{Verlinde:2006bc} the decay rate was estimated by considering the bubble nucleation rate sourced by the D5-brane walls. This leads to a formula of the general form $\Gamma_{\mathrm{mono}} \sim \exp\left(-\frac{27 \pi^2}{2} \frac{\mathcal{T}^{4}}{\mathcal{E}^3}\right)$ where $\mathcal{T}$ is proportional to the tension of the wall obtained from a D5-brane wrapped on the 3-chain, and $\mathcal{E}$ is an energy density set by the (stringy) volume of the 2-cycle.} In terms of the radial slicing of the conical geometry $X$, we can depict this as a locally increasing volume $\mathrm{Vol}(\alpha)$ which eventually drops to zero size once a ``maximal size'' is reached.\footnote{In addition to the decay channels considered in \cite{Kasai:2015exa}, one could also imagine that the geometric moduli of $Y$ itself might dynamically adjust to ``shorten'' the size of the 3-chain. Such an effect can be suppressed provided we have already stabilized complex structure moduli using various fluxes. This can indeed be arranged (see e.g., \cite{Gukov:1999ya, Giddings:2001yu}).}

From the perspective of the 4D spacetime, the dynamics of this eventual annihilation involves an initial contraction, and then expansion of the 3-ball with the monopole at its core. Outside the expanding bubble, the $U(1)^{(1)}_{\mathrm{mag}}$ is unbroken, but inside it is broken, signaling the presence of confinement of $U(1)^{(0)}_{\mathrm{elec}}$ degrees of freedom (see figure \ref{fig:MonoDefect}).

\begin{figure}
    \centering
    \scalebox{1}{
    \begin{tikzpicture}
	\begin{pgfonlayer}{nodelayer}
		\node [style=none] (0) at (-3, 0.75) {};
		\node [style=none] (1) at (-3.8, -0.1) {};
		\node [style=none] (2) at (-3, -0.75) {};
		\node [style=none] (3) at (-2.25, 0) {};
		\node [style=ShadedCircRed2] (4) at (-3, 0) {};
		\node [style=none] (10) at (-0.5, 0) {};
		\node [style=none] (11) at (5.5, 0) {};
		\node [style=none] (12) at (-3, 0) {};
		\node [style=ShadedCircRed] (14) at (3, 0) {};
		\node [style=none] (15) at (3, 0) {};
		\node [style=none] (16) at (3, 0.75) {};
		\node [style=none] (17) at (2.25, 0) {};
		\node [style=none] (18) at (3, -0.75) {};
		\node [style=none] (19) at (4.375, 0) {};
		\node [style=ShadedCirc] (20) at (3, 0) {};
		\node [style=ShadedCircRed3] (21) at (3, 0) {};
		\node [style=none] (22) at (-3, 1.125) {\color{blue} $\mathcal{U}_{\eta}(\Sigma_2)$};
		\node [style=none] (23) at (5.25, -1.5) {\color{blue} $\mathcal{U}_{\eta}(\Sigma_2)$};
		\node [style=none] (24) at (-3, -0.425) {\small $\Gamma_Y$};
		\node [style=none] (25) at (3, -1.75) {\small $\Gamma_Y$};
		\node [style=none] (26) at (-0.5, -0.375) {$\alpha_Y$};
		\node [style=none] (28) at (5.5, -0.375) {$\alpha_Y$};
		\node [style=none] (29) at (-3, 2.25) {$G_N\neq 0, t \ll \tau_{\textnormal{mono}} $};
		\node [style=none] (30) at (3, 2.25) {$G_N\neq 0, t\gg \tau_{\textnormal{mono}} $};
		\node [style=none] (31) at (3, 0.8) {\small $F2(\Sigma_2\times I)$};
		\node [style=none] (32) at (4.5, -1.25) {};
		\node [style=none] (33) at (3.5, -0.5) {};
        \node [style=none] (34) at (-3.8, 0.1) {};
	\end{pgfonlayer}
	\begin{pgfonlayer}{edgelayer}
		\draw [style=ThickLine] (12.center) to (10.center);
		\draw [style=ThickLine] (19.center) to (11.center);
		\draw [style=ArrowLineRight] (32.center) to (33.center);
        \draw [style=BlueDottedLight, bend left=87, looseness=1.75] (1.center) to (3.center);
		\draw [style=BlueDottedLight, bend left=87, looseness=1.75] (3.center) to (34.center);
	\end{pgfonlayer}
\end{tikzpicture}
    }
    \caption{Left: The monopole is linked by the symmetry operator $\mathcal{U}_{\eta}(\Sigma_2)$. Right: After unwinding along $\Gamma_Y$ the monopole has grown beyond $\Sigma_2$ (dotted blue circle). When the flux 4-brane passes through the D5-brane, the flux 2-brane $F2$ associated with NSNS 3-form flux is created. It localizes on $\Sigma_2\times I$ with boundaries on the D5-brane and the flux 4-brane, here $\Sigma_2\times I$ is the difference of two spatial 3-balls. }
    \label{fig:Flux2}
\end{figure}
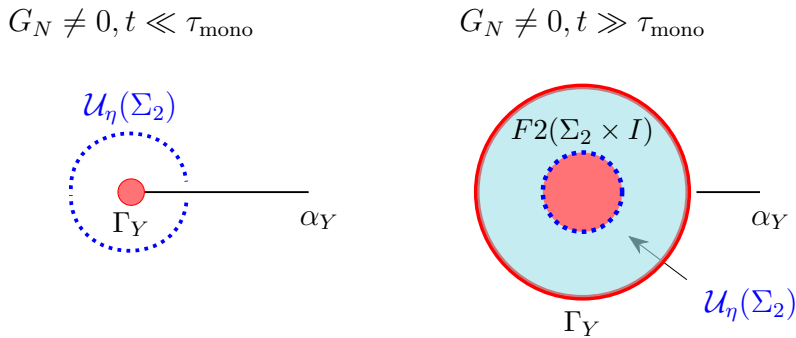

This is all in accord with expectations from generalized symmetries. First of all, in the global compactification geometry, the fluxbrane supported ``at infinity'' on $\Lambda_3$ will in general now collapse to the tip of the cone. In this collapsing process, the fluxbrane used to detect the monopole will also pass into the core of the monopole, in accord with our discussion of subsection \ref{ssec:FINITE}. The collapse is not entirely trivial, since in the process of collapsing, a TFT supported on a flux 2-brane will be created via the Hanany-Witten brane creation effect. See figure \ref{fig:Flux2} for a depiction.

\section{Summary and Future Directions}

In this note we have presented a general proposal for how to engineer topological symmetry operators for continuous symmetries via branes ``wrapped at infinity''. The end result can be summarized as interpreting the lift of the WZ action of a $p$-brane in differential cohomology as an ``ordinary'' $(p+2)$-form, integrated against the worldvolume of a flux $(p+1)$-brane. We have used this implementation to revisit the phase structure of Verlinde's metastable monopole, and in particular, have shown that such constructions are also sensitive to finite size effects in monopole configurations. In the remainder of this section we discuss a few natural generalizations.

We have mainly focused on the specific case of Verlinde's metastable monopole, where the interior region exhibits full confinement of electric degrees of freedom. One could also entertain an intermediate situation where $U(1)^{(0)}_{\textnormal{elec}}$ only partially confines, leaving a subgroup $\mathbb{Z}_n\subset U(1)^{(1)}_{\textnormal{mag}}$ unbroken. This occurs whenever the 2-cycle $\alpha$ does not completely trivialize, rather $n$ copies of $\alpha$ trivialize as a 2-cycle in $Y$. Said differently, there only exists a 3-chain $\Gamma$ such that $\partial \Gamma=n\alpha$ and $\alpha$ is a degree $n$ torsion cycle in $H_2(Y)$. Constructing explicit compact geometries which realize this phenomenon would be quite interesting.

In the context of string constructions, it was recently found that certain higher-group structures can also be detected by suitable defects (see e.g., \cite{Cvetic:2022imb, DelZotto:2022joo}). These analyses focused on QFTs with discrete higher-form symmetries, but one could in principle look for examples with continuous higher-form symmetries, where there are sometimes non-trivial constraints (see e.g., \cite{Genolini:2022mpi}). It would be interesting to revisit these questions from the perspective of the topological operators generated by fluxbranes.

Especially once gravity is switched on, it is natural to consider the fate of the metastable monopole when it is thrown into a black hole. Tracking the ultimate fate of the monopole, fluxbrane, and black hole in this setting would likely be very instructive.


\section*{Acknowledgments}

MC is supported in part by the Slovenian Research Agency (ARRS No. P1-0306)
and Fay R. and Eugene L. Langberg Endowed Chair funds.
The work of MC, JJH, and ET is supported by DOE (HEP) Award DE-SC0013528.
The work of MC, JJH, and MH is supported in
part by a University Research Foundation grant at the University of Pennsylvania. The
work of MC and MH is also supported by the Simons Foundation Collaboration grant
\#724069 on ``Special Holonomy in Geometry, Analysis and Physics''.

\appendix

\section{Fluxbranes from Brane / Anti-Brane Pairs}\label{app:fluxfrombranes}
In this Appendix we show how flux $(p+1)$-branes can be realized as soliton backgrounds of D-brane / anti-D-brane pairs in Type II string theories. Consider first the universal WZ Lagrangian on D-branes \cite{Minasian:1997mm}, viewed as a $(p+1)$-form:
\begin{equation}
    \mathcal{L}^{(p+1)}_{\textnormal{WZ}}=C\wedge e^{(F_2-B_2)} \wedge \sqrt{\frac{\mathcal{A}(R_T)}{\mathcal{A}(R_N)}}
\end{equation}
where the pullback of the bulk RR forms to the brane are given by the formal sum $C=\sum_{p+1}C_{p+1}$ with $p+1$ even for IIB and $p+1$ odd for IIA, and $\mathcal{A}({R_{T}}),\mathcal{A}({R_{N}})$ denotes the A-roof genus for the tangent and normal bundles. Recall from line \eqref{eq:symop} that the action for a flux $(p+1)$-brane with worldvolume $Y_{p+2}$ is
\begin{equation}\label{eq:wzfluxbrane}
    2\pi i \eta \int_{Y_{p+2}} \mathcal{F}^{(p+2)}_{\textnormal{WZ}}
\end{equation}
where locally we have $\mathcal{F}^{(p+2)} = d \mathcal{L}^{(p+1)}_{\mathrm{WZ}}$. The parameter
$\eta\in \mathbb{R}/\mathbb{Z}$ characterizes its flux in the source equation
\be
* F_{p+2}=\eta \delta_{Y_{p+2}}.
\ee
We find that when $\eta$ is rational $\eta\in \mathbb{Q}/\mathbb{Z}$, one can always realize the flux $(p+1)$-brane from the following brane / anti-brane system
\be
\begin{aligned}
    &\left( D(p+2)+\overline{D(p+2)} \right)+\\&\left( N_1\times D(p+4)+ N_1\times \overline{D(p+4)} \right)+\\
    &\left( N_2\times D(p+6)+ N_2\times \overline{D(p+6)} \right)+\dots
    \end{aligned}
\ee
where the worldvolumes of all branes / anti-brane pairs of the same dimension coincide, and they are nested, e.g.:
\be
X_{p+3}\subset X_{p+5}\subset \dots\,.
\ee

On $X_{p+3}$, i.e., the worldvolume of the $D(p+2)$ and $\overline{D(p+2)}$, the flux $(p+1)$-brane is sourced by a monodromy for the $U(1)$ gauge field $A$ on the system\footnote{More specifically, on the brane / anti-brane system there is a $U(1)\times U(1)$ gauge theory (along with tachyonic bifundamentals) and we will always mean the $(+1,-1)$ diagonal combination of these $U(1)$'s.} which is localized if we take the singular connection
\begin{equation}\label{eq:monofluxbrane}
    A=\eta\delta_{Y_{p+2}}
\end{equation}
where $\delta_{Y_{p+2}}$ is oriented normal to $Y_{p+2}\subset X_{p+3}$ and therefore a 1-form on $X_{p+3}$. However this does not produce the correct coefficients for the terms in \eqref{eq:wzfluxbrane} which is why the other brane/anti-brane pairs are necessary. Along $X_{p+5}$, i.e., the worldvolume of the $D(p+4)$ and $\overline{D(p+4)}$, the flux $(p+1)$-brane is characterized by a localized Chern-Simons density
\begin{equation}\label{eq:csfluxbrane}
    CS_3(A)=\eta_1\delta_{Y_{p+2}}
\end{equation}
where similar to above $A$ is a connection for diagonal $U(N_1)$ in $U(N_1)\times U(N_1)$ gauge theory on $X_{p+5}$. So as to not create a flux $(p+2)$-brane, we must take a solution \eqref{eq:csfluxbrane} such that $\mathrm{Tr}A=0$ which implies
\begin{equation}
    \eta_1=\frac{m}{N_1}
\end{equation}
for some $m\in \mathbb{Z}$ since Chern-Simons integrals associated to the Lie algebra $\mathfrak{su}(N_1)$ can only take such values. If $(p+2)>7$ and $\eta \in \mathbb{Q}/\mathbb{Z}$ then we can tune $N_1$ and $\eta_1$ to reach the WZ action of the flux $(p+1)$-brane \eqref{eq:wzfluxbrane}, while if $(p+2)\leq 7$ then we continue on to produce higher-dimensional brane / anti-brane pairs as indicated above.

A final comment is that since we dealing with the topological limit of a brane / anti-brane system, there is a natural formulation available in terms of connections on supergroups. See \cite{Vafa:2001qf} for additional discussion on this point.

\bibliographystyle{utphys}
\bibliography{MetaMono}

\end{document}